\documentclass[aps,prl,twocolumn,amsfonts,amssymb,amsmath,floats,floatfix,showpacs,preprintnumbers,superscriptaddress]{revtex4-1}

\usepackage{graphicx}
\usepackage{color}
\usepackage{dcolumn}
\usepackage{bm}
\usepackage{hyperref}
\usepackage{relsize}

\hypersetup{
colorlinks=true,
urlcolor= blue,
citecolor=blue,
linkcolor= blue,
bookmarks=true,
bookmarksopen=false,
}

\begin{document}

\title{Polarity-Driven Surface Metallicity in SmB$_6$}

\author{Z.-H. Zhu}
\affiliation{Department of Physics {\rm {\&}} Astronomy, University of British Columbia, Vancouver, British Columbia V6T\,1Z1, Canada}
\author{A. Nicolaou}
\affiliation{Department of Physics {\rm {\&}} Astronomy, University of British Columbia, Vancouver, British Columbia V6T\,1Z1, Canada}
\affiliation{Quantum Matter Institute, University of British Columbia, Vancouver, British Columbia V6T\,1Z4, Canada}
\author{G. Levy}
\affiliation{Department of Physics {\rm {\&}} Astronomy, University of British Columbia, Vancouver, British Columbia V6T\,1Z1, Canada}
\affiliation{Quantum Matter Institute, University of British Columbia, Vancouver, British Columbia V6T\,1Z4, Canada}
\author{N.\,P. Butch}
\affiliation{CNAM, Department of Physics, University of Maryland, College Park, Maryland 20742, USA}
\author{P. Syers}
\affiliation{CNAM, Department of Physics, University of Maryland, College Park, Maryland 20742, USA}
\author{X.\,F. Wang}
\affiliation{CNAM, Department of Physics, University of Maryland, College Park, Maryland 20742, USA}
\author{\\J. Paglione}
\affiliation{CNAM, Department of Physics, University of Maryland, College Park, Maryland 20742, USA}
\author{G.\,A. Sawatzky}
\affiliation{Department of Physics {\rm {\&}} Astronomy, University of British Columbia, Vancouver, British Columbia V6T\,1Z1, Canada}
\affiliation{Quantum Matter Institute, University of British Columbia, Vancouver, British Columbia V6T\,1Z4, Canada}
\author{I.\,S. Elfimov}
\email{elfimov@physics.ubc.ca}
\affiliation{Department of Physics {\rm {\&}} Astronomy, University of British Columbia, Vancouver, British Columbia V6T\,1Z1, Canada}
\affiliation{Quantum Matter Institute, University of British Columbia, Vancouver, British Columbia V6T\,1Z4, Canada}
\author{A. Damascelli}
\email{damascelli@physics.ubc.ca}
\affiliation{Department of Physics {\rm {\&}} Astronomy, University of British Columbia, Vancouver, British Columbia V6T\,1Z1, Canada}
\affiliation{Quantum Matter Institute, University of British Columbia, Vancouver, British Columbia V6T\,1Z4, Canada}

\date{\today}

\begin{abstract}
By a combined angle-resolved photoemission spectroscopy and density functional theory study, we discover that the surface metallicity is polarity-driven in SmB$_6$.\,Two surface states, not accounted for by the bulk band structure, are reproduced by slab calculations for coexisting B$_6$ and Sm surface terminations. Our analysis reveals that a metallic surface state stems from an unusual property, generic to the (001) termination of all hexaborides: the presence of boron $2p$ dangling bonds, on a polar surface. The discovery of polarity-driven surface metallicity sheds new light on the 40-year old conundrum of the low-temperature residual conductivity of SmB$_6$, and raises a fundamental question in the field of topological Kondo insulators regarding the interplay between polarity and nontrivial topological properties.
\end{abstract}

\pacs{71.20.-b, 73.20.-r, 79.60.-i, 71.27.+a}

\maketitle

Highly-renormalized $f$-electrons are the quasiparticles underlying heavy-fermion behavior \cite{Coleman:book}. When conduction electrons interact with these atomically-confined $f$-electrons, in dynamically screening their magnetic moment, the quasiparticle spectrum is modified by the opening of a narrow charge gap at low temperatures and a Kondo-insulating state is realized \cite{Fisk:review, Ueda:RMP, Riseborough}, as in the archetypical case of FeSi \cite{Fisk:review,Schlesinger:FeSi,Damascelli:FeSi}. If, in addition, spin-orbit coupling is larger than the many-body Kondo gap, topological surface states (TSS) \cite{Fu:2007, Hasan:2010PRM, Qi:TIrmp, Zhu:TSS} are predicted to exist, defining a new class of strongly-correlated electron systems: the  {\it topological Kondo insulators} (TKI) \cite{Dzero:TKI, Dzero:TKIprb}. As for possible TKI candidates, SmB$_6$ has initially been suggested  \cite{Dzero:TKI}. Later, model calculations based on density-functional theory (DFT) predicte three Dirac surface states (SS) residing at the time-invariant points in the surface Brillouin zone \cite{Takimoto:JPSJ, Lu:dftPRL}. Interestingly, SmB$_6$ has long been known for its anomalous resistivity behavior at low temperatures \cite{SmB6:first, Allen:mixvalent, Cooley:kondo}: it undergoes a metal-to-Kondo-insulator-like transition below 50\,K, with an exponential increase of 4 orders of magnitude from 15\,K to 5\,K and saturation at lower temperatures \cite{Cooley:kondo}.
\begin{figure}[b!]
\includegraphics[width=1\linewidth]{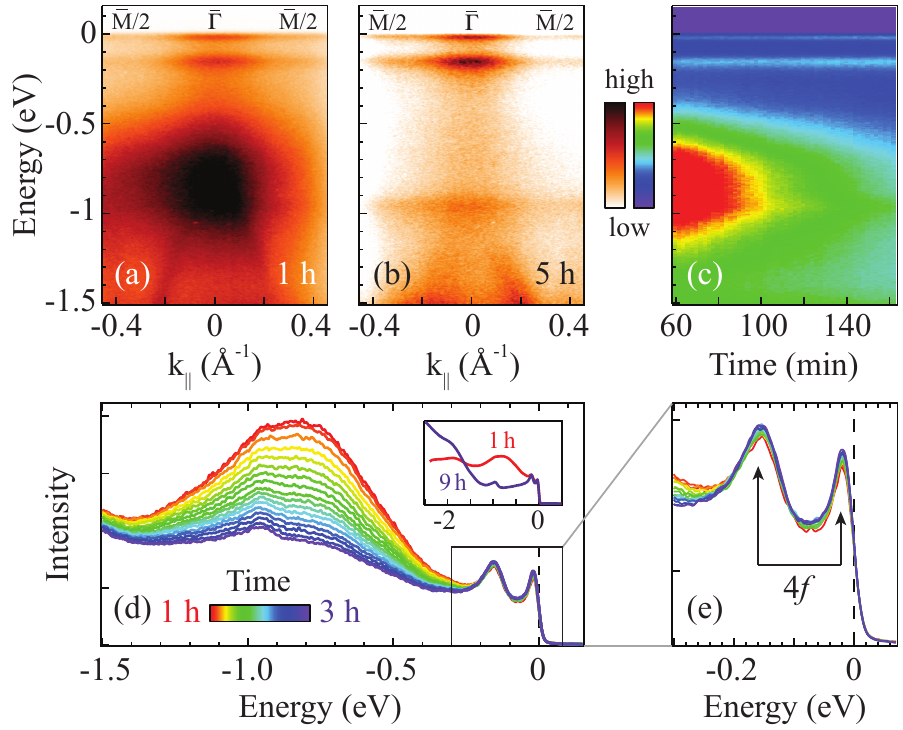}
\caption {\label{fig:timeARPES} (color online). (a),(b) Time evolution of the ARPES dispersion of SmB$_6$ measured along $\mathrm{\bar M}\!-\!\bar\Gamma\!-\!\mathrm{\bar M}$ at $5\times 10^{-11}$ torr and $T\!=\!6$\,K: (a) 1 h and (b) 5 h after cleaving. (c)--(e) Time evolution of the $k$-integrated ARPES maps: continuous sequence of 2-min-averaged data in (c), with only 1 curve out of 3 shown in (d,e). Despite the strong dynamics, corresponding to a transfer of spectral weight from low to high energies [inset of (d)], the Sm 4$f$-multiplets are remarkably stable (e).}
\end{figure}
The residual conductivity below 5\,K was attributed to in-gap states \cite{Gorshunov:ingap, Sluchanko:ingap}, but their nature has remained mysterious for the past 40 years. The prediction of TKI behavior might provide a long-sought-after solution, in the form of a TSS within the Kondo gap.

A surface origin for the low-temperature conductivity of SmB$_6$ was indicated by recent transport studies \cite{Wolgast:R, JP:pcs, Kim:hall,Kim:impurity,Thomas:WAL}; angle-resolved photoemission spectroscopy (ARPES) \cite{Miyazaki:ARPES, Xu:ARPES, Neupane:ARPES, Jiang:ARPES,Golden:arpes} and quantum oscillations \cite{Li:QS} also provided tentative evidence for the existence of two-dimensional SS.
\begin{figure*}[t!]
\includegraphics[width=0.95\linewidth]{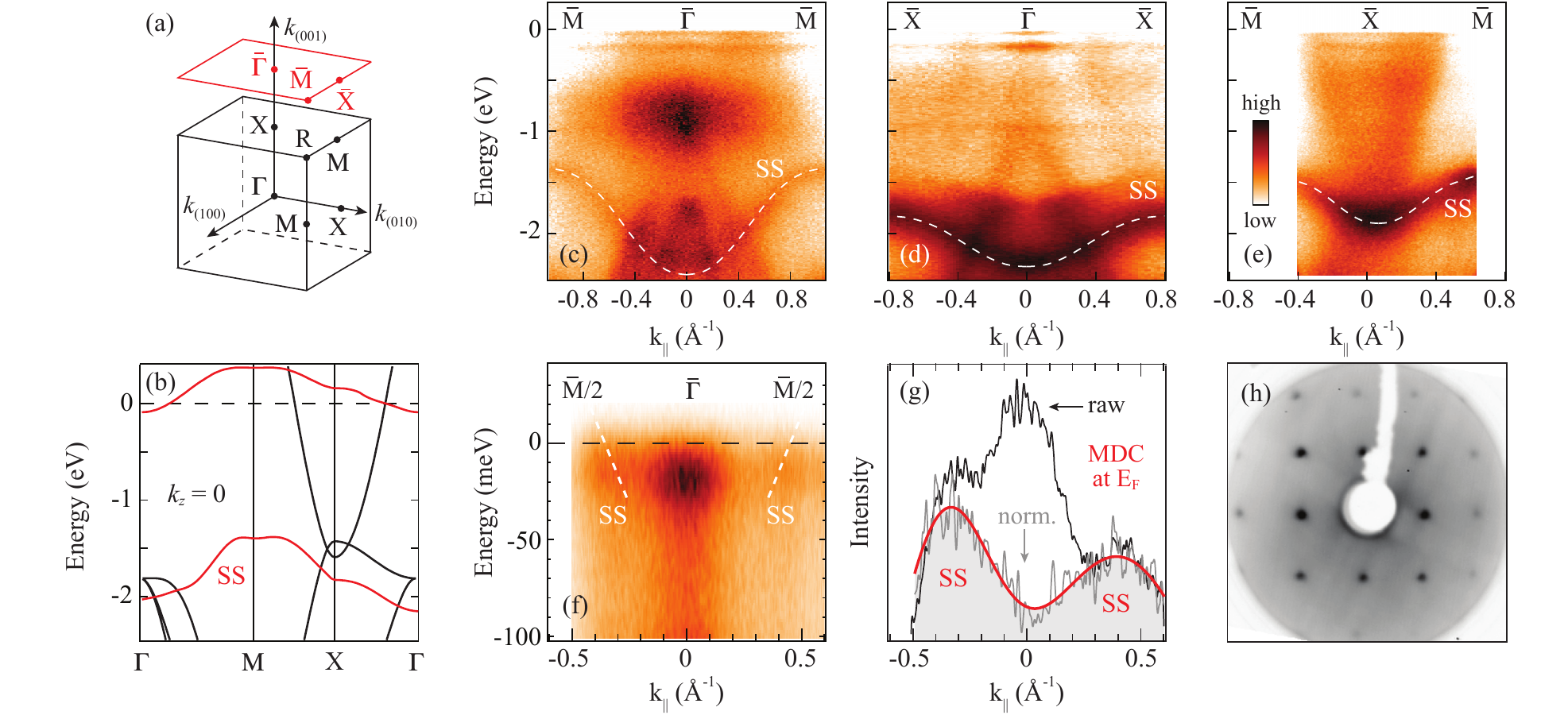}
\caption {\label{fig:ARPES} (color online). (a) SmB$_6$ bulk and projected (001) surface Brillouin zones.  (b) SmB$_6$ bulk band structure at $k_z\!=\!0$ (black; $f$ bands removed for clarity), with in addition the B-2$p$ dangling-bond-derived surface states (SS, red), as revealed by our combined ARPES and slab-DFT study. (c)--(e) ARPES dispersions along $\mathrm{\bar M}\!-\!\bar\Gamma\!-\!\mathrm{\bar M}$ (c),  $\mathrm{\bar X}\!-\!\bar\Gamma\!-\!\mathrm{\bar X}$ (d), and  $\mathrm{\bar M}\!-\!\mathrm{\bar X}\!-\!\mathrm{\bar M}$ (e) measured at 6\,K with 21.2\,eV photons, corresponding to $k_z \approx 0$ \cite{Golden:arpes}. (f) Enlarged dispersion along $\mathrm{\bar M}\!-\!\bar\Gamma\!-\!\mathrm{\bar M}$, where no bulk bands are predicted to cross $E_F$ (b); a pile-up of intensity at $E_F$ -- evidenced by a 3 peak profile in the raw MDC at $E_F$ in (g), which reduces to a 2 peak structure when the raw data in (f) are normalized to compensate for the cross-section enhancement at $\bar\Gamma$ -- provides evidence for the existence of a surface electron pocket around the $\bar\Gamma$ point.  In (c--f) the white-dashed lines highlight the observed B-2$p$ SS. Note that the spectra in (c) are measured on a fresh cleave, while all other data are from stabilized surfaces, including the LEED pattern in (h).}
\end{figure*}
The key outstanding question is whether these are chiral topological states or instead induced by other mechanisms. In fact, a clear-cut demonstration of topological invariance is still lacking and -- most critical -- SS associated with boron dangling bonds are well known to exist in the hexaborides. These SS are often located at about $-$2\,eV binding energy or slightly above the Fermi level ($E_F$) \cite{Monnier:LaB6dft,Trenary:review}, and are thus believed to be generally nonmetallic. Note, however, that the (001) surface of hexaborides is polar, enabling the partial filling of the unoccupied SS via a small chemical potential shift, as also suggested for LaB6 by an early inverse-photoemission study \cite{Morimoto:IPHE}. If crossing $E_F$, such a SS would provide an alternative mechanism for surface metallicity, raising important questions regarding the role of TSS and the potential interaction between polarity- and topology-driven SS.

Here, studying SmB$_6$ by ARPES and DFT slab calculations \cite{supplementary}, we demonstrate the existence of polarity-driven surface metallicity in SmB$_6$. Two sets of SS observed by ARPES -- not accounted for by the bulk band structure -- are well reproduced by DFT calculations performed for a slab geometry with both B$_6$ and Sm terminations (consistent with the lack of a natural cleavage plane). We show that while B-2$p$ dangling-bond-derived SS are present on both terminations, a SS electron pocket forms at the $\bar\Gamma$ point -- as a result of the polarity-induced electronic reconstruction -- only on the B$_6$-terminated (001) surface. Our observations reveal the presence of a polarity-driven SS distinct from the predicted TSS in SmB$_6$.

\begin{figure}[t!]
\includegraphics[width=0.95\linewidth]{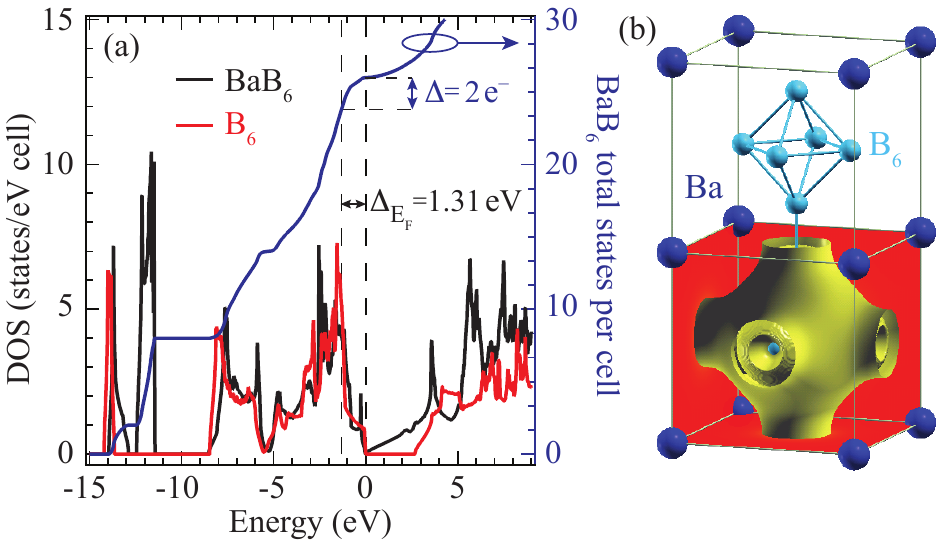}
\caption {\label{fig:DOS} (color online). (a) Comparison between the DOS of BaB$_6$ and a model material made of B$_6$ only  (note that the differences seen above $E_F$ and at $-$12\,eV stem from the Ba $5d$ and $5p$ states, respectively, which, however, play no role in the discussion of hexaborides' ionicity). The overall DOS profiles agree with each other after a 1.31\,eV  upward $E_F$ shift for B$_6$, corresponding to the addition of 2 electrons as indicated by the integration of the BaB$_6$ DOS (blue). (b) Crystal structure of BaB$_6$ with one unit cell showing the surface of constant charge-density at the isovalue 0.025 e/$\mathrm{\AA}^{3}$.}
\end{figure}

\begin{figure*}[t!]
\includegraphics[width=0.97\linewidth]{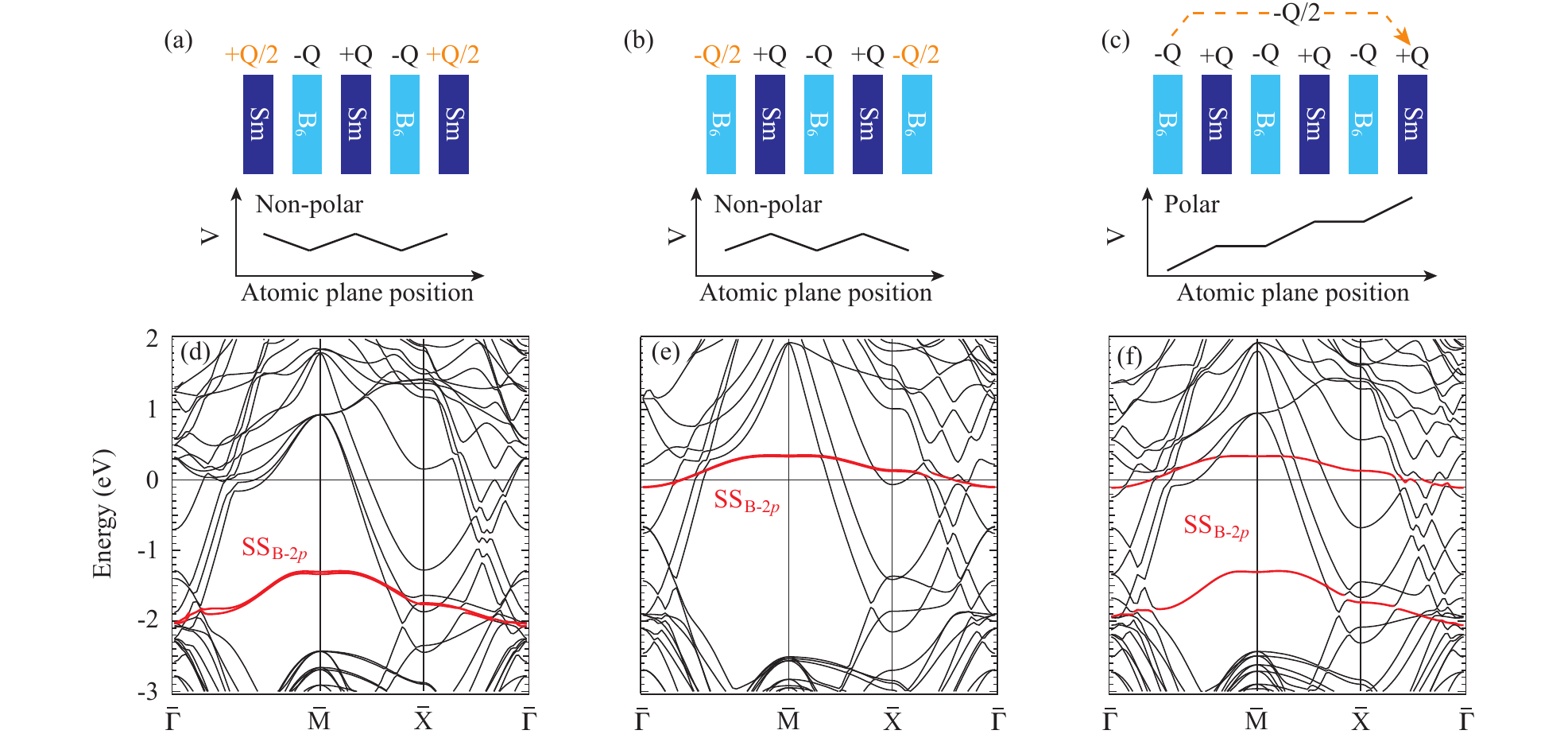}
\caption {\label{fig:SlabBand} (color online). (a,b) Symmetric,
nonstoichiometric slabs -- either Sm (a) or B$_6$ (b) terminated on both sides -- are nonpolar. (c) The nonsymmetric, stoichiometric
slab -- with one Sm and one B$_6$ termination -- is instead polar, because of the divergent electrostatic potential V associated with the
alternating $-\mathrm{Q}$/$+\mathrm{Q}$ charged planes, leading to {\it electronic reconstruction}: the transfer of $\mathrm{Q}$/2
electronic charge from the B$_6$ to the Sm termination, resulting in a charge stacking analogous to (a),(b). (d--f) Slab band structures
calculated for the geometries in (a--c), with a 5-unit-cell thickness. Two sets of SS -- {\it one metallic} -- are derived from the B $2p$ dangling bonds (red bands). The near-$E_F$ black bands are Sm $5d$-states, which in the slabs span the bulk $k_z$ dispersion; these do not contribute to the low-temperature conductivity due to the opening of the Kondo gap, consistent with scanning tunnelling spectroscopy \cite{Yee:stm}  ($f$-bands and related $5d$-$4f$ hybridization are here not accounted for, since the Sm $4f$-states are treated as core levels).}
\end{figure*}

Let us start by pointing out a peculiarity of the ARPES results from cleaved SmB$_6$, likely important also for transport studies performed on annealed surfaces. Generally, ARPES spectra become  progressively broader with time because of aging of the as-cleaved surfaces \cite{Christian:Ru214}. However, in SmB$_6$ -- a material without a natural cleavage plane and whose cleaved (001) surface thus presents Sm- and B$_6$-terminations with equal probability -- we observe an opposite dynamics, even at temperatures as low as 6\,K. The freshly cleaved samples always exhibit an intense broad feature around $-\!0.8$\,eV [Fig.\,\ref{fig:timeARPES}(a)], coexisting with the nondispersive $4f$-multiplets at $-$0.02, $-$0.15, and $-$0.97\,eV \cite{Denlinger2000716}. Surprisingly, this broad structure is progressively suppressed with time, eventually disappearing few hours after cleaving [Fig.\,\ref{fig:timeARPES}(b)], as shown in detail by the time evolution of the $k$-integrated ARPES maps in Figs.\,\ref{fig:timeARPES}(c--e). Note that -- since the Sm $4f$-states are utterly unaffected -- this dynamics must be associated with the self-annealing of the as-cleaved surface and related SS (more later, in light of the results in Fig.\,\ref{fig:ARPES}--\ref{fig:SlabBand}).

The band structure of SmB$_6$ in a 2.5\,eV binding-energy window, as determined by ARPES, is presented in Fig.\,\ref{fig:ARPES}. We note that low-energy electron diffraction (LEED) on our SmB$_6$ (001) cleaved surfaces shows a clear $1\!\times\!1$ structure [Fig.\,\ref{fig:ARPES}(h)], indicating a predominantly structurally-unreconstructed surface. The high-symmetry-direction ARPES dispersion in Fig.\,\ref{fig:ARPES}(c--e) can be compared to DFT bulk calculations \cite{Lu:dftPRL, Antonov:dft}, here presented in black in Fig.\,\ref{fig:ARPES}(b) for $k_z\!=\!0$ (with Sm $4f$-states removed for clarity): we observe a qualitative correspondence for the large Sm-5$d$ electron Fermi pocket at $X$ [Fig.\,\ref{fig:ARPES}(e)], and the valence bands around $-\!2$\,eV at $\Gamma$  [Fig.\,\ref{fig:ARPES}(c)]. 
Note however that there are also ARPES features not expected in the bulk band structure of Fig.\,\ref{fig:ARPES}(b). The most obvious one is the band seen at all momenta around $-\!1.8$\,eV binding energy with $\sim$1\,eV bandwidth, highlighted by a white dashed line in Figs.\,\ref{fig:ARPES}(c--e). In analogy with the results obtained on the (001) surface of LaB$_6$, CeB$_6$, PrB$_6$, and NdB$_6$ \cite{Trenary:review}, it might be attributed to subsurface-boron dangling bonds from metal-terminated regions (i.e., La, Ce, Pr, Nd, and here Sm). In addition, while for $k_z\!\simeq\!0$ we also do not expect any bulk bands crossing $E_F$ along $\mathrm{\bar M}-\bar\Gamma-\mathrm{\bar M}$, the corresponding image plot [Fig.\,\ref{fig:ARPES}(f)] shows a clear pile-up of intensity resulting in a three-peak structure in the momentum distribution curve (MDC) at $E_F$ [Fig.\,\ref{fig:ARPES}(g)]. We note, however, that an ARPES cross-section enhancement is observed around $\bar\Gamma$ at all energies [Fig.\,\ref{fig:ARPES}(c-f)], which might mask the location of the true Fermi crossings. Thus, to uncover the latter, in Fig.\,\ref{fig:ARPES}(g) we are also showing  the `normalized' MDC -- i.e., obtained after normalization of the ARPES intensity map in Fig.\,\ref{fig:ARPES}(f) to the peak height of each $k$-resolved EDC. This is effectively equivalent to plotting the EDC leading-edge-midpoint dispersion, and reveals the presence of an electron pocket centered at the $\bar\Gamma$ point [Figs.\,\ref{fig:ARPES}(f,g)].

The detection of a SS at $E_F$ around $\bar\Gamma$ is consistent with other ARPES studies \cite{Xu:ARPES,  Neupane:ARPES, Jiang:ARPES,Golden:arpes}. As for the report of a second metallic SS at $\bar X$ \cite{Xu:ARPES,  Neupane:ARPES, Jiang:ARPES}, possibly connected to the predicted TSS \cite{Takimoto:JPSJ, Lu:dftPRL}, this was shown to stem from the hybridization between Sm $d$-band and $-\!0.02$\,eV $f$-state  \cite{Golden:arpes}, and thus belongs to the bulk electronic structure also visible in our data. We also note that no temperature dependence is seen in our data at $E_F$ beyond conventional Fermi function broadening; in addition, while the detection of TSS at $E_F$ is challenging due to stringent resolution requirements, the Dirac cone predicted around $-40$\,meV at $\bar{X}$ -- thus in a region of $k$-space free of bulk bands (see Fig.\,5 in Ref.\,\onlinecite{Lu:dftPRL}) -- should be observable, but is here also not detected. As we will argue below based on our DFT slab analysis --\,and anticipated in Fig.\,\ref{fig:ARPES}(b) that combines bulk and surface bands\,-- all the  bulk-unexpected states are SS derived from B-2$p$ dangling-bonds. Most important, an {\it otherwise unoccupied SS} is pushed below $E_F$ at $\bar\Gamma$ by electronic reconstruction, leading to a {\it polarity-driven surface metallicity}.

Before examining its effects, we address if the prerequisite for a polar catastrophe is satisfied in hexaborides: an ionic nature of the material, giving rise to a stack of alternating planes of opposite charge. Because DFT cannot properly treat the correlation effects of the $f$-states, and these are not relevant to this discussion, to demonstrate the ionic nature of hexaborides we choose for a simplicity BaB$_6$ -- a material with a band structure similar to that of SmB$_6$ but without $f$-states. Fig.\,\ref{fig:DOS}(a) shows a comparison between the BaB$_6$ density of states (DOS) and that of an artificial material containing only B$_6$ octahedra; the latter is rigidly shifted in energy to compensate for 2 missing electrons.
Evidently, the DOS profiles are very similar in both shape and peak positions, suggesting that the cation (here Ba) simply donates 2 electrons to the B$_6$ sublattice, leaving its electronic structure essentially unperturbed. Thus what emerges for the hexaborides is an ionic picture in which, in striking contrast to transition metal oxides, the hybridization between cation and ligand orbitals is not a key factor in band formation. 
This is also illustrated by the charge-density plot in Fig.\,\ref{fig:DOS}(b), where the charge bounded only around the B$_6$ octahedron highlights: (i) the {\it ionic nature} -- with an alternation of oppositely-charged planes along the [001] direction, responsible for the polar instability; (ii) the {\it covalent bonding} within the B$_6$ network -- which necessarily leads to the formation of boron dangling bonds at the surface.

Having established their ionic nature, we examine the response of the (001) surface to a polar catastrophe. Hexaborides crystallize in the CsCl crystal structure, consisting of two interpenetrating cubic lattices [Fig.\,\ref{fig:DOS}(b)]. Along the [001] direction, one can think of it as a stack of alternating planes of opposite charge, separated by a half lattice constant [Fig.\,\ref{fig:SlabBand}(c)]. Because of the monotonic increase of electrostatic potential with thickness, leading to a diverging surface energy,  an ideal termination of such series cannot exist unless it is stabilized by substantial structural, chemical, and/or electronic reconstruction \cite{Tasker, Hesper:polar}. In hexaborides, a rather unique feature is that a purely electronic reconstruction might be favoured by the presence of the half-filled B$_6$ dangling bonds, on both metal- and boron-terminated surfaces. 

So far, DFT studies of the SS in hexaborides have only been performed for La-terminated LaB$_6$ slabs \cite{Monnier:LaB6dft,Trenary:review} -- as motivated by the mostly metal-terminated surfaces obtained by polishing and annealing -- and a metallic SS was not found. These results, however, do not fully capture the case of as-cleaved hexaborides, where metal (La, Sm, etc.) and B$_6$ terminations coexist with equal probability. To this end, we start from two artificially symmetric slabs, with either Sm or B$_6$ terminations; their nonstoichiometric nature leads to a reduction of the outermost-plane charge to half of the bulk value -- i.e., the polar catastrophe is fixed by construction [Figs.\,\ref{fig:SlabBand}(a,b)]. Note that the Sm 4$f$-states are here treated as core level for simplicity \cite{supplementary}, since their interaction with the B$_6$ dangling bonds is negligible, as seen in Fig.\,\ref{fig:timeARPES}(d,e) and previously reported \cite{patil:092106}. As shown in Fig.\,\ref{fig:SlabBand}(d), the Sm-terminated slab possesses a SS of B-2$p$ character at $-$2\,eV binding energy, in close agreement with the ARPES data in Figs.\,\ref{fig:ARPES}(c--e). Most importantly, the B$_6$-terminated slab shows a  B-2$p$ SS crossing $E_F$ [Fig.\,\ref{fig:SlabBand}(e)], consistent with the electron pocket observed at $\bar\Gamma$ in Figs.\,\ref{fig:ARPES}(f,g) -- and also the inverse photoemission results from LaB$_6$ \cite{Morimoto:IPHE}.

To capture the actual situation of as-cleaved SmB$_6$, i.e. a polar system with both Sm and B$_6$ terminations, the calculations is repeated for the asymmetric, stoichiometric slab of Fig.\,\ref{fig:SlabBand}(c); the self-consistent DFT solution in Fig.\,\ref{fig:SlabBand}(f) is analogous to the combination of the results from the two symmetric slabs [Figs.\,\ref{fig:SlabBand}(d,e)]. This comparison allows also determining the driving mechanism -- structural versus electronic -- behind the surface metallicity. In fact, the excellent agreement between the band structure results for both polar and nonpolar slabs, and the observation of the same structural relaxation in the DFT calculations \cite{relax}, indicate that structural effects in proximity of the surface are neither a consequence of -- nor a solution for -- the polar surface instability. The latter is stabilized through an electronic reconstruction rather than a rearrangement of the surface atomic structure, consistent also with the 1$\times$1 diffraction pattern measured by LEED indicating a predominately structurally unreconstructed surface [Fig.\,\ref{fig:ARPES}(h)].

Finally, we can now understand also the ARPES intensity dynamics seen in  Fig.\,\ref{fig:timeARPES}. Because of the lack of a natural cleavage plane, the as-cleaved surface might exhibit a disordered distribution of B$_6$-like molecules and Sm atoms; to minimize its total energy, this might slowly relax to form large Sm- and B$_6$-terminated domains. Correspondingly, as illustrated by the inset of Fig.\,\ref{fig:timeARPES}(d) and the comparison of Fig.\,\ref{fig:ARPES}(c) and \ref{fig:ARPES}(d), spectral weight is transferred from the broad feature at $-\!0.8$\,eV to the  dispersing SS of the B$_6$ and Sm terminations, centered at about $+\!0.2$\,eV and $-\!1.8$\,eV respectively (Fig.\,\ref{fig:SlabBand}). Energetically, the spectral weight of the disordered surface -- with its random Sm-B$_6$ coordination, statistically in between that of ideal Sm and B$_6$ terminations -- should be located half-way between the SS from the Sm and B$_6$ domains: i.e. at $-\!0.8$\,eV, as indeed experimentally observed.

In conclusion, by a combined ARPES and DFT analysis we have shown the existence of a metallic SS in SmB$_6$, which is associated with an intrinsic, general property of hexaborides: the presence of boron dangling bonds, on a polar surface. The discovery of polarity-driven surface metallicity sheds new light on the 40-year old conundrum of the low-temperature residual conductivity of SmB$_6$. In addition, our study raises a fundamental question in the TKI field regarding how the polarity of a surface would affect its nontrivial topological properties. The possible interplay between these two distinct types of SS requires further investigation, both in theory and experiment. 

We acknowledge M. M. Yee, J. E. Hoffman, S. Wirth, S. Kirchner, and L. H. Tjeng for helpful discussions. This work was supported by the Max Planck - UBC Centre for Quantum Materials, the Killam, Alfred P. Sloan, Alexander von Humboldt, and NSERC's Steacie Memorial Fellowships (A.D.), the Canada Research Chairs Program (A.D.), NSERC, CFI, and CIFAR Quantum Materials. 

\vspace{-0.3cm}
\bibliography{SmB6_revised}

\end{document}